\documentstyle[preprint,prb,aps,epsf]{revtex}
\begin{document}
\tightenlines
\draft

\title{Segregation and charge-density-wave order in the
spinless Falicov-Kimball model}
\author{J. K. Freericks$^{\dagger}$ and R. Lema\'nski$^*$}
\address{$^{\dagger}$Department of Physics, Georgetown University, 
Washington, DC 20057 USA\\
$^*$Institute of Low Temperatures and Structure Research, Polish Academy of
Sciences, Wroc\l aw, Poland}
\maketitle
\begin{abstract}
The spinless Falicov-Kimball model is solved exactly in the limit of 
infinite-dimensions on both the hypercubic and Bethe lattices.  The competition
between segregation, which is present for large $U$, and charge-density-wave
order, which is prevalent at moderate $U$, is examined in detail.  We
find a rich phase diagram which displays both of these phases.  The model
also shows nonanalytic behavior in the charge-density-wave transition 
temperature when $U$ is large enough to generate a correlation-induced gap
in the single-particle density of states.
\end{abstract}
\widetext
\section{Introduction}

The Falicov-Kimball model\cite{falicov_kimball} is the simplest Fermionic model 
for crystallization\cite{lieb_kennedy}, where the system has a phase transition
from a disordered (liquid) phase at high temperature to an ordered (solid)
phase as the temperature is lowered. It similarly can be viewed as a binary
alloy problem where the presence of an ion indicates an A species and the
absence of an ion is a B species. 

In this model itinerant (spinless) electrons interact with static
ions through an on-site Coulomb interaction.  Many-body effects enter via
the statistical mechanics associated with annealed averaging.
It is the simplest many-body problem that can be solved exactly in the limit of 
large dimensions\cite{metzner_vollhardt}.

Brandt and Mielsch\cite{brandt_mielsch}
presented the first solution of this problem using dynamical mean-field theory.
Their solution illustrated how a period-two charge-density-wave phase is
stabilized at low temperatures.  Freericks\cite{freericks} later
showed that the model also illustrated incommensurate charge-density-wave
order and phase separation. That work concentrated on the case where the
ions were half filled on a hypercubic lattice.  Segregation was favored
at large interaction strength, and incommensurate order disappeared when
the interaction strength became larger than the order of the hopping matrix 
element.

Recent work\cite{letfulov,freericks_macris_gruber} has shown that the
segregation principle\cite{freericks_falicov} holds in the infinite-dimensional
limit---as $T$ is lowered the system undergoes a phase transition that
separates it into electron-rich and ion-rich regions (when the interaction
energy becomes infinite).  This result, coupled with the rigorous proof
of segregation in one-dimension\cite{lemberger} and approximate results in 
two-dimensions\cite{gruber_jedrzejewski_lemberger,watson_lemanski}, provides 
compelling evidence
for segregation to hold in all dimensions.  We offer no proof of that statement
here.  Instead, we just want to comment that such a result is in the same
spirit as
the Brandt-Schmidt\cite{brandt_schmidt} and Lieb-Kennedy\cite{lieb_kennedy}
result that when the electron and ion concentrations are equal to one-half, the
system orders in a period-two ordered phase for all dimensions (and has a 
finite-temperature phase transition for $d\ge 2$).  This tendency toward 
phase separation could be the mechanism that drives strongly correlated systems
like the cuprates or the nickelates towards charge-stripe formation, where the
stripes arise from a minimization of the free energy when both the tendency
toward phase separation and the long-range Coulomb interaction are taken into
account\cite{kivelson_emery}.  Further results about the Falicov-Kimball 
model can be found in a recently completed review\cite{gruber_review}.

In this contribution we examine what happens to the spinless Falicov-Kimball
model as the interaction strength is made finite and the system engages in
a competition between phase separation (segregation) and charge-density-wave
order.  We find a number of interesting results for the phase diagrams that
differ from what occurred in the infinite-interaction-strength 
limit\cite{freericks_macris_gruber}.

This manuscript is organized as follows: Section II describes the formalism,
Section III presents the results for both the Bethe lattice and the hypercubic
lattice, and Section IV presents our conclusions.

\section{Formalism}

The spinless Falicov-Kimball model is represented by the following
Hamiltonian:

\begin{equation}
H=-\frac{t^*}{2\sqrt{d}}\sum_{<i,j>}c_i^{\dagger}c_j+E\sum_iw_i
+U\sum_ic_i^{\dagger}c_iw_i,
\label{eq: hamiltonian}
\end{equation}
where $c_i^{\dagger}$ ($c_i$) creates (destroys) a conduction electron at
lattice site $i$ and $w_i=0$ or 1 is a classical variable that measures
the number of ions at lattice site $i$.  The hopping matrix connects
nearest neighbors $i$ and $j$ and has magnitude $t^*/(2\sqrt{d})$ which
scales inversely as the square root of the dimensionality $d$.  We choose
$t^*=1$ to be our energy scale.  $E$ is the site energy for the ions and $U$ 
is the on-site Coulomb interaction between electrons and ions.  For simplicity
we will consider the case of positive $U$, since negative $U$ can be mapped
onto this case with a particle-hole transformation\cite{lieb_kennedy}.

In the thermodynamic limit, the local lattice Green's function is defined to be
\begin{equation}
G_n=G(i\omega_n)=-\int_0^{\beta}d\tau e^{i\omega_n\tau}\frac{{\rm Tr}\langle 
e^{-\beta (H-\mu N)}
T_{\tau} c(\tau)c^{\dagger}(0)\rangle}{{\rm Tr}
\langle e^{-\beta (H-\mu N)}\rangle},
\label{eq: greendef}
\end{equation}
where $i\omega_n=i\pi T(2n+1)$ is the Fermionic Matsubara frequency,
$\beta=1/T$ is the inverse temperature, $N$ is a number of conduction electrons,
and $T_{\tau}$ denotes $\tau$-ordering. A chemical potential $\mu$ is used to 
set the electron concentration $\rho_e=\langle c^{\dagger}c\rangle$ and the site 
energy $E$ is adjusted
to yield the ion concentration $\rho_i=\langle w\rangle$. The angle brackets in
Eq.~(\ref{eq: greendef}) denote the sum over ionic configurations.
The local Green's function is determined by mapping onto an atomic problem
in a time-dependent field, with the following action
\begin{equation}
S_{at}=\int_0^{\beta}d\tau\int_0^{\beta}d\tau^{\prime}c^{\dagger}(\tau)
G_0^{-1}(\tau-\tau^{\prime})c(\tau^{\prime})+U\int_0^{\beta}d\tau
c^{\dagger}(\tau)c(\tau)w+Ew,
\label{eq: action}
\end{equation}
where $w=0, 1$ is the ion number for the atomic site
and $G_0^{-1}$ is the mean-field or effective-medium
Green's function, which is determined self-consistently (as described below).
The atomic Green's function, with the action in Eq.~(\ref{eq: action}),
is computed to be
\begin{equation}
G_n=\frac{1-\rho_i}{G_0^{-1}(i\omega_n)}+\frac{\rho_i}{G_0^{-1}(i\omega_n)-U},
\label{eq: greenatomic}
\end{equation}
and the local lattice Green's function satisfies
\begin{equation}
G_n=\int_{-\infty}^{\infty}d\epsilon \frac{\rho(\epsilon)}{i\omega_n+\mu-
\Sigma_n-\epsilon},
\label{eq: greenlocal}
\end{equation}
where $\rho(\epsilon)$ is the noninteracting density of states for the
infinite lattice and $\Sigma_n$ is the self-energy.  The self-consistency 
relation is that the self-energy $\Sigma_n$ in Eq.~(\ref{eq: greenlocal})
must coincide with the self-energy of the atomic problem, i.~e.
\begin{equation}
\Sigma(i\omega_n)=G_0^{-1}(i\omega_n)-G_n^{-1}.
\label{eq: sigmadyson}
\end{equation}
Equations (\ref{eq: greenatomic}), (\ref{eq: greenlocal}), and
(\ref{eq: sigmadyson}) constitute the dynamical mean-field theory for 
homogeneous phases.  In the limit $d\rightarrow\infty$ Eq.~(\ref{eq: 
sigmadyson}) is an exact equation for the lattice problem. 

These equations must be solved numerically for the general case (an analytic 
simplification\cite{freericks_macris_gruber} occurs on the Bethe lattice
when $U=\infty$).
We use Jarrell's iterative algorithm\cite{jarrell} to solve this problem:
(i) begin with the self-energy set equal to zero $\Sigma_n=0$; (ii) use
Eq.~(\ref{eq: greenlocal}) to determine the local Green's function;
(iii) solve for the effective medium by employing Eq.~(\ref{eq: sigmadyson});
(iv) find the new local Green's function from Eq.~(\ref{eq: greenatomic});
(v) extract a new self energy from Eq.~(\ref{eq: sigmadyson}); and (vi)
repeat steps (ii-v) until the self-energy does not change from one iteration
to the next.  We use a relative error of one part in $10^7$ as our convergence
criterion.  These equations rapidly converge in most cases, but occasionally
require damping of oscillations to force them to converge, rather than enter a
periodic limit cycle.  These equations can also be solved on the real axis,
where the Matsubara frequency is replaced by the real frequency
($i\omega_n\rightarrow \omega+i\delta$).

At high temperatures the system is in a homogeneous phase, with a uniform
charge density. As $T$ is lowered, the system can undergo a phase transition
to a charge density wave. The temperature below which the homogenous phase is 
unstable is found by calculating the divergence of the relevant susceptibility. 
The detailed formulas for these susceptibilities have appeared 
elsewhere\cite{brandt_mielsch,freericks,freericks_zlatic} and will
not be repeated here.  We do not consider incommensurate order in this
contribution.  Incommensurate order is easily handled in the hypercubic
lattice\cite{brandt_mielsch,freericks}, but is problematic on the Bethe
lattice, where it appears that higher-order periodic phases always have
first-order phase transitions\cite{bethe_incomm}.
We investigate two different possibilities here:
(i) the two-sublattice chessboard charge-density-wave phase, which
has different electron and ion charge densities on the two sublattices of the
bipartite lattice, and (ii) the segregated phase, where the system separates 
into two uniform phases with different electron and ion densities. The former is
the $(\pi,\pi,\pi,...)$ susceptibility and the latter is the uniform
susceptibility.

We also need to calculate the Helmholz free energy for these systems
in order to perform a Maxwell construction to track the first-order
phase transition to the segregated phase (the transition to the chessboard phase
is always continous). This free energy can be
expressed either as a summation over Matsubara frequencies, as first shown
by Brandt and Mielsch\cite{brandt_mielsch}, or it can be expressed as
an integral over the interacting density of states, as first shown by 
Ramirez, Falicov, and Kimball\cite{falicov_kimball}.  We choose the former
form, since there is no analytic form for the interacting density of states
when $U$ is finite.  Hence, the free energy becomes
\begin{eqnarray}
F(\rho_e,\rho_i)=-T\ln2+T\rho_i\ln\rho_i+T(1-\rho_i)\ln(1-\rho_i)+\mu(\rho_e-
{1\over 2})+{U\rho_i\over 2} \\  \nonumber
+T\int d\epsilon\rho(\epsilon)\sum_n\ln[{G_0(i\omega_n)i\omega_n\over 
(i\omega_n+\mu-\Sigma_n-\epsilon)G_n}]-T\rho_i\sum_n\ln[1-UG_0(i\omega_n)].
\end{eqnarray}
One must be careful in evaluating this expression, since the integrand, which
involves the summation of a logarithm over the Matsubara frequencies, requires
a large frequency cutoff to converge.

The Maxwell construction for the phase-separated (segregated)
state consists of taking a 
weighted average of the free energy in two homogeneous phases with
densities $(\rho_e^{(1)},\rho_i^{(1)})$ and 
$(\rho_e^{(2)},\rho_i^{(2)})$ subject to the system having the correct
{\it average} electron and ion concentration.  In equations, we take
\begin{equation}
F_{seg}(\rho_e,\rho_i)=\alpha 
F(\rho_e^{(1)},\rho_i^{(1)})+(1-\alpha)F(\rho_e^{(2)},\rho_i^{(2)})
\end{equation}
where
\begin{equation}
\label{rhoe}
\rho_e=\alpha \rho_e^{(1)}+(1-\alpha)\rho_e^{(2)},
\end{equation}
\begin{equation}
\label{rhoi}
\rho_i=\alpha \rho_i^{(1)}+(1-\alpha)\rho_i^{(2)} 
\end{equation}
The electron concentrations are determined by setting a common chemical
potential between the two phases and the constraint of Eq.~(\ref{rhoe}).
Of the six parameters ($\alpha$, $\mu$, $\rho_e^{(1)}$, $\rho_e^{(2)}$,
$\rho_i^{(1)}$, and $\rho_i^{(2)}$) needed
to specify the segregated phase, only two are independent variables.  We 
use the ion concentrations $\rho_i^{(1)}$ and $\rho_i^{(2)}$ as our
independent variables.  Our minimization procedure is identical to the one
used in the infinite-$U$ case\cite{freericks_macris_gruber}: (i) we first
choose a coarse grid for both $\rho_i^{(1)}$ and $\rho_i^{(2)}$ and compute
the average free energy for all points on that grid, and locate the minimum;
(ii) the ion density $\rho_i^{(1)}$ is fixed at this coarse minimum, and
$\rho_i^{(2)}$ is varied over a fine grid to find the new minimum;
(iii) the ion density $\rho_i^{(2)}$ is fixed at this new minimum, and
$\rho_i^{(1)}$ is varied over a fine grid to find the new minimum; and
(iv) both ion densities are varied over a final fine grid centered at the 
approximate minimum to complete the minimization procedure.  We find that the
minima rarely change in step (iv) which illustrates the convergence of this 
method.  This multistep convergence procedure is much more efficient than
just minimizing over the fine grid from the start.

\section{Results}

We perform calculations for two different lattices---the 
infinite-coordination-number Bethe lattice where 
$\rho(\epsilon)=\sqrt{4-\epsilon^2}/(2\pi)$ and the infinite-dimensional
hypercubic lattice where $\rho(\epsilon)=\exp(-\epsilon^2)/\sqrt{\pi}$.
In general, computations for the Bethe lattice are simpler than for the 
hypercubic lattice, because many of the integrals over the density of
states can be performed analytically.  But we find that there is 
little difference between the results for the two lattices, as can be seen
in the results presented below.

We begin by showing the transition temperatures to the chessboard 
(two-sublattice) charge density wave and the spinodal-decomposition
temperature for segregation, as determined by finding the temperature
where the relevant susceptibility diverges.  Figure 1 displays the results
for the case where $\rho_i=0.2$ on the Bethe lattice for two different values
of $U$.  In the weak-coupling regime, there is no competition between the
chessboard phase and segregation, because the two regions do not overlap, but
when $U$ is made larger, one can see an overlap between these regions.
It may appear then that there are regions in the weak-coupling regime
where the homogenous phase is 
stable all the way down to $T=0$, but we believe that this will not
be the case in general.  As seen in the work on the hypercubic lattice
at $\rho_i=0.5$, the region that appeared to be a homogeneous phase 
turned out to be one that displayed incommensurate order\cite{freericks}.  
We expect a similar result to take place here, but due to the difficulty in
calculating incommensurate order on the Bethe lattice (which is typically
a first-order transition) we have not investigated that question here.

Furthermore, a kink appears in the $T_c(\rho_e)$ curve for the chessboard
phase.  This kink occurs at the filling $\rho_e=1-\rho_i$, which is a special
filling for the spinless Falicov-Kimball model.  When $U$ is large enough,
this is the filling where the system undergoes a metal-insulator transition
(in this case with $\rho_i=0.2$
the critical value of $U$ is approximately 1.86).
The interacting density of states for the electrons generates a gap, and 
increasing the electron
filling from just below $1-\rho_i$ to just above $1-\rho_i$
results in a large shift in the electronic chemical potential as it moves
from the lower to the upper band\cite{leinung_vandongen}.  
What is remarkable is that this
metal-insulator transition illustrates itself via a kink in the chessboard
phase transition temperature!  A similar result can be seen in the 
Hubbard model, but was not pointed out in the original 
paper\cite{jarrell_freericks}. Beyond $U=3$ the antiferromagnetic transition
temperature curves display the same kink at half filling as seen in the
Falicov-Kimball model, and the presence of such kinks appears to be another
way to infer that the system has a gap in the single-particle density of
states, which does not require performing calculations on the real axis
(which are much more difficult for quantum Monte Carlo simulations).

In Figure 2 we show the spinodal-decomposition temperature for segregation
on the Bethe lattice for both $\rho_i=0.5$ and $\rho_i=0.2$.  Notice how
the two pieces of the phase diagram move towards each other to meet at
$\rho_e=1-\rho_i$.  We also include the result for $U=\infty$ which has a 
mirror symmetry about the line $\rho_e=(1-\rho_i)/2$.  That symmetry is
absent for finite-$U$ and develops slowly as $U$ increases.  There are
no states at finite energy with $\rho_e>1-\rho_i$ when $U=\infty$ so only
one branch is included in the spinodal-decomposition temperature.

The transition temperature to the chessboard phase on the Bethe lattice is
shown in Figure 3 for the same cases $\rho_i=0.5$ and $\rho_i=0.2$.  Here we
separate the figures into those at weak coupling (a) and (c), where there is 
no kink in the phase diagram, and those at strong coupling
(b) and (d), where a kink is present because
of the gap in the single-particle density of states (for $\rho_i=0.5$ this
occurs at $U=2$).  The chessboard phase is 
stable only in a narrow window around $\rho_e=1-\rho_i$ when $U$ is large,
but migrates towards $\rho_e=0.5$ for smaller values of $U$. The competition
between segregation and the chessboard phase is the strongest when 
$\rho_e\approx 1-\rho_i$ and $U$ is around two times of the bandwidth.
Notice how the case with $\rho_i=0.5$ displays an additional reflection
symmetry about $\rho_e=0.5$.  This particle-hole symmetry disappears,
and the phase diagrams possess a strong asymmetry when $\rho_i\ne 0.5$.

Figures 4 and 5 display the identical results as Figures 2 and 3 respectively, 
but this time are plotted for the hypercubic lattice rather than the Bethe
lattice.  It is remarkable how similar the results are for these two lattices
(with the exception of an overall scale factor).  The kinks in the
chessboard phase diagram appear to be sharper on the hypercubic lattice, but
otherwise the results are nearly identical with each other.  Due to the
similarity of the results for the Bethe and hypercubic lattices, we have
chosen to concentrate only on the computationally simpler Bethe lattice for the
free-energy analysis (we have verified the similarity of the free-energy
phase diagrams for the hypercubic and Bethe lattices for a few cases).

A Maxwell construction is needed to calculate the phase
diagram when the system phase separates.  Just like the case where 
$U=\infty$, we find that the phase diagram has special homogeneous densities 
$(\rho_e^*,\rho_i^*)$ where the first-order phase transition (binodal, $T_b$)
and the spinodal-decomposition temperature (spinodal, $T_s$)
coincide. The point corresponds to a critical 
temperature $T_c=T_{s,~\rm max}$) where for a given $U$ both the first-order 
transition temperature and the spinodal-decomposition temperature share 
a maximum. At this point both of the electron densities approach 
the homogeneous density $(\rho_e^{(1)}\rightarrow\rho_e^*$ and 
$\rho_e^{(2)}\rightarrow\rho_e^*$) as the temperature approaches the 
transition temperature from below (and likewise for $\rho_i$). Obviously the 
transition is continuous at this point.  In the general case, we find only one
of the two pairs tends towards the homogeneous values of the fillings
as $T_b$ is approached, and the phase transition is discontinuous (in this case
we find $\alpha$ approaches either 0 or 1 as $T_b$ is approached).
The phase diagrams are complicated three-dimensional curves in $\rho_i$,
$\rho_e$, and $T$ space.  We project those curves onto different planes in order
to summarize our results.

Figure 6 contains the projection of the segregation phase diagram onto the
$\rho_e$--$\rho_i$ plane for the Bethe lattice.  The diamonds indicate the
values of the electron and ion concentrations at the maximum of the
spinodal-decomposition temperature for a given value of $U$.  These maxima
are monotonic in $\rho_e$, but are nonmonotonic in $\rho_i$ increasing from
0.58 at $U=0.25$ to 0.8 for $U=2$ and then decreasing to 0.65 as
$U\rightarrow\infty$.  The solid curves display the pairs of densities
that the system phase separates into as a function of temperature when in
the segregated phase.  As $T\rightarrow 0$ we find all systems go to the
states with $\rho_e^{(1)}=0$ and $\rho_i^{(1)}=1$ and $\rho_e^{(2)}=
\rho_e^*/(1-\rho_i^*)$ and $\rho_i^{(2)}=0$.  The dashed lines are straight
lines that connect these two points, and are guides to the eye.  The 
chain-dashed line is a similar plot for the case where $U=\infty$.  Note
how these solid curves are nearly straight lines for both small and large $U$,
and how they become curved only for cases of intermediate $U$.  The maximal
spinodal-decomposition temperature does monotonically increase with $U$ as
shown in Figure 7. 

It is remarkable that the results we obtained in the infinite-dimension limit as
$T\rightarrow 0$ are similar to those found in the ground-state in one 
dimension\cite{lach_lyzwa_jedrzejewski,gruber_ueltschi_jedrzejewski,gajek_jedrzejewski_lemanski} 
and two dimensions\cite{gruber_jedrzejewski_lemberger,watson_lemanski}. 
In all these cases (for $U>0$) the segregated phase characterized by the pair of
densities $(\rho_e, \rho_i)$ is a mixture of the fully occupied phase (where the
ions clump together) without electrons, i.e. $(\rho_e^{(1)},\rho_i^{(1)})=(0,1)$
and the empty phase (without ions) with a finite density of electrons equal to 
$\rho_e/(1-\rho_i)$, i.e. $(\rho_e^{(2)},\rho_i^{(2)})=(\rho_e/(1-\rho_i),0)$. 
The two regions where the segregated phase is stable consist of those points of 
the ($\rho_e, \rho_i$) plane that satisfy one of the inequalities: 
$0<\rho_e<(1-\rho_i) b_d(U)$ or $(1-\rho_i ) b_d(U)<\rho_e<1$, where $b_d(U)$ 
[$b_d(U)>0$] is an increasing function of $U$ tending towards unity when $U$ 
goes to infinity ($d$ denotes the spatial dimension). In the one-dimensional
case a transcendental equation for $b_1(U)$ has been 
derived\cite{gruber_ueltschi_jedrzejewski}.

With an increase of $U$, the stability of the segregated phase for
the one and two-dimensional phase diagrams spreads over the whole 
region of densities $\rho_e$ and $\rho_i$ except for the 
unit-density case $\rho_e+\rho_i=1$, where periodic
phases are stable. In one dimension, the unit-density
phase corresponds to the most homogenous distribution of the ions for any $U$. 
In two dimensions,
the ions are also arranged periodically (in the unit-density case) but 
their arrangement changes with $U$ (there is no unique ``most homogenous 
distribution'' in two dimensions).

If $\rho_i=0.5$ and the unit-density condition is fulfilled ($\rho_i+\rho_e=1$),
then the charge-ordered phase is found to be stable in the infinite-dimensional
limit 
(in fact, this is also for a region of $\rho_e$  close to that given by the 
unit-density condition). However our calculations show this property is relevant
for moderate $U$ only. Presumably the order of the limits 
$d\rightarrow \infty$; $U\rightarrow \infty$ and $\rho_e\rightarrow 1-\rho_i$ 
must be taken properly to get the charge-density order in 
this case. We cannot compare the one and two-dimensional results for 
$\rho_i=0.2$ 
($\rho_e=1-\rho_i=0.8$) with those in the infinite $U$ limit because we 
rectricted ourselves (for technical reasons)
to the chessboard-type charge-ordering only. 

For finite $U$, the rest of the $(\rho_e, \rho_i$) region (apart from the 
areas occupied by the segregated and the unit-density phases) of the one and  
two-dimensional phase diagrams contain a number of
charge-density-wave phases that differ from the chessboard one 
(as well as their mixtures). We expect that the similar effect will occur in the
infinite-dimensional limit for intermediate densities, where  
the homogenous phase appeared to be stable down to zero temperature (see
Fig.~1), but where we expect incommensurate order to prevail. 

Figures 8 and 9 show the projection of the phase diagram onto the
$\rho_e$--$T$ and $\rho_i$--$T$ planes respectively.  At any given temperature,
a horizontal line intersects a solid line of
the phase diagram at two points, corresponding
to the pair $(\rho_e^{(1)},\rho_e^{(2)})$ and $(\rho_i^{(1)},\rho_i^{(2)})$
respectively.  The solid lines are the binodal (first-order) phase-transition
lines, and the dashed lines are the spinodal (second-order) phase-transition
lines where the system becomes locally unstable.
Although these phase diagrams appear to have similar shapes
to those seen at $U=\infty$, we were unable to determine any kind of
appropriate scaling form which could collapse the data onto an universal 
scaling form.

An example of the general case, where the phase transition is discontinuous,
is shown in Figure 10.  Here the spinodal and binodal transition temperatures
are not equal to each other for a given pair of densities ($\rho_e, \rho_i$).  
We have chosen the
case of $\rho_e=0.15$, $\rho_i=0.5$, and $U=4$, and show results only for the
$\rho_i$--$T$ plane. Note that one of the ion densities has a discontinous 
jump at $T_b$ whereas the other one changes smoothly when the
temperature is lowered
below $T_b$. This occurs because of the nucleation of the new phase inside 
the (old) high-temperature phase. 

\section{Conclusions}

The main result of this work is the pervasiveness of phase separation and
the segregation principle in the Falicov-Kimball model in infinite-dimensions.
We see that it survives for all values of $U$, and that it can take up a 
large portion of the phase space in the system.  In addition, the transition
temperatures become larger as $U$ grows, and the phase-separated state takes
over the entire phase diagram except possibly the point where $\rho_e=1-\rho_i$.
Since this is precisely the result seen in the one-dimensional\cite{lemberger} 
and two-dimensional\cite{gruber_jedrzejewski_lemberger,watson_lemanski} cases, this result strongly suggests
that the phenomenon of segregation is indeed independent of 
dimensionality.  Such a general principle should have a fundamental physical
reason that drives its behavior,
and this begs for a general proof that would hold in arbitrary
dimensions.  We offer no such proof here, since we are unable to determine what
this general principle is.  In one dimension, segregation is driven by a 
lowering of the kinetic energy by placing all electrons in as large a ``box'' 
as possible.  This kinetic-energy-driven effect should hold in all dimensions,
but the analysis is much more complicated for $d>1$.  We do believe that this
general principle is important in the phenomena of stripes, since it must 
contribute to the ability of a system like the Hubbard model to form stripes.

We also find that there are some regions where this segregation can compete
with charge-density-wave order.  These regions are fairly small in the 
phase diagram, since they occur near $\rho_e=1-\rho_i$ for moderate values of 
$U$.  In this region there can also be competition between incommensurate
order (which we have not considered due to it's technical difficulties
on the Bethe lattice) and either phase separation or chessboard 
charge-density-wave order.

Finally, we discovered an interesting slope discontinuity (nonanalyticity)
in the chessboard-phase transition temperature which occurs when the
single-particle density of states generates a correlation-induced gap.  Such
a signature of a correlation-induced gap is ubiquitous, and it can also
be seen in the Hubbard model when it is beyond the Mott transition.  While
we believe the formation of an anomalous kink in the phase diagram implies
the generation of a correlation-induced gap, we once again offer no proof,
and simply state that such an observation will shed insight on 
metal-insulator transitions, but it is not a substitute for calculations of
the single-particle density of states.

\begin{figure}[t]
\centerline{
\epsfxsize=3.5in \epsffile{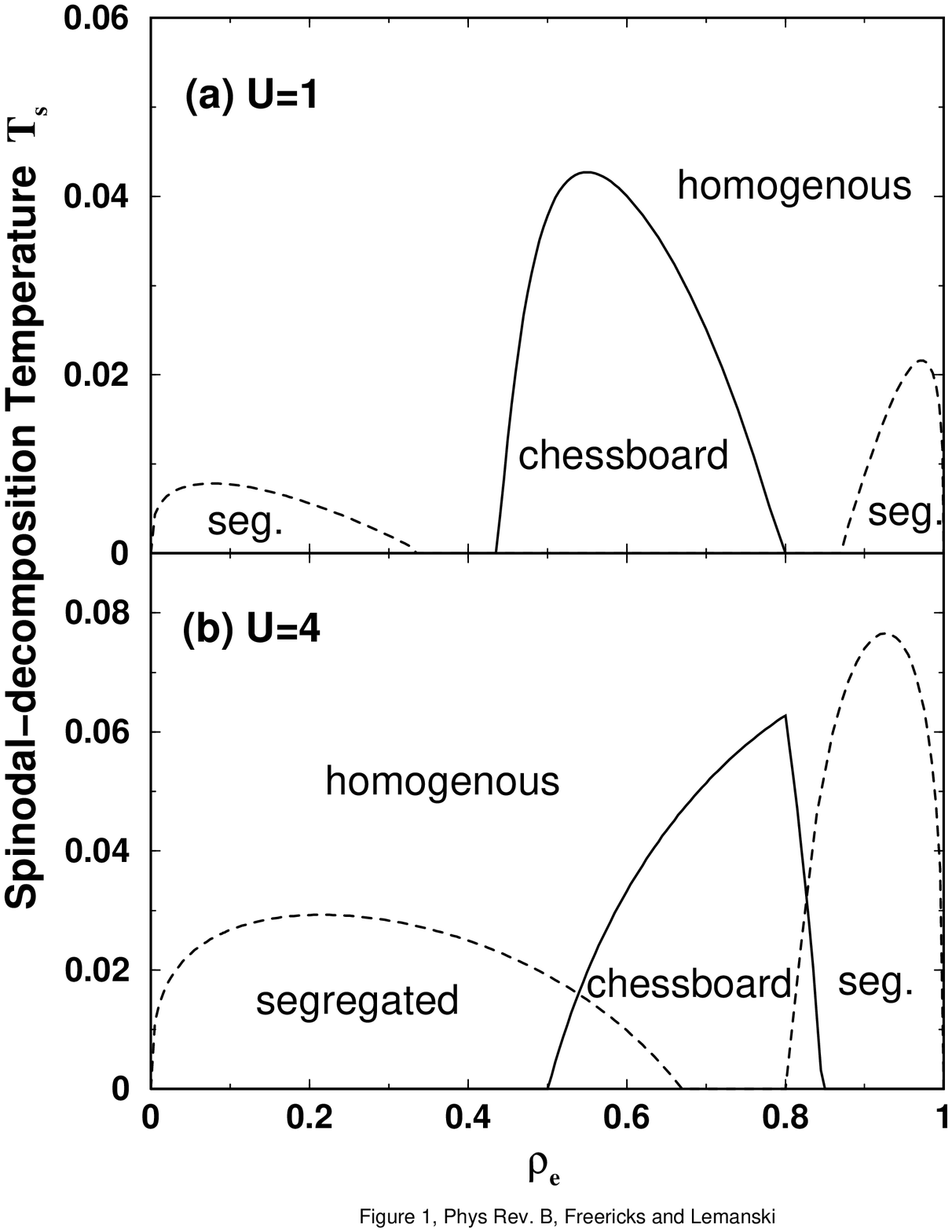}}
\caption{Phase diagrams to the chessboard charge-density-wave phase and
the spinodal-decomposition temperature for the segregated phase on the
Bethe lattice with $\rho_i=0.2$.  Figure 1(a) is the case $U=1$ where the
chessboard phase and the segregated phase do not compete with each other.
Figure 1(b) is the case with $U=4$ where there is an overlap of the
spinodal phase lines, indicating a competition between segregation and charge
density wave formation,
and where a well-developed kink can be seen in the chessboard phase
diagram, which occurs due to a correlation-induced gap in the
single-particle density of states, as described in the text.}
\end{figure}

\begin{figure}[t]
\centerline{
\epsfxsize=3.5in \epsffile{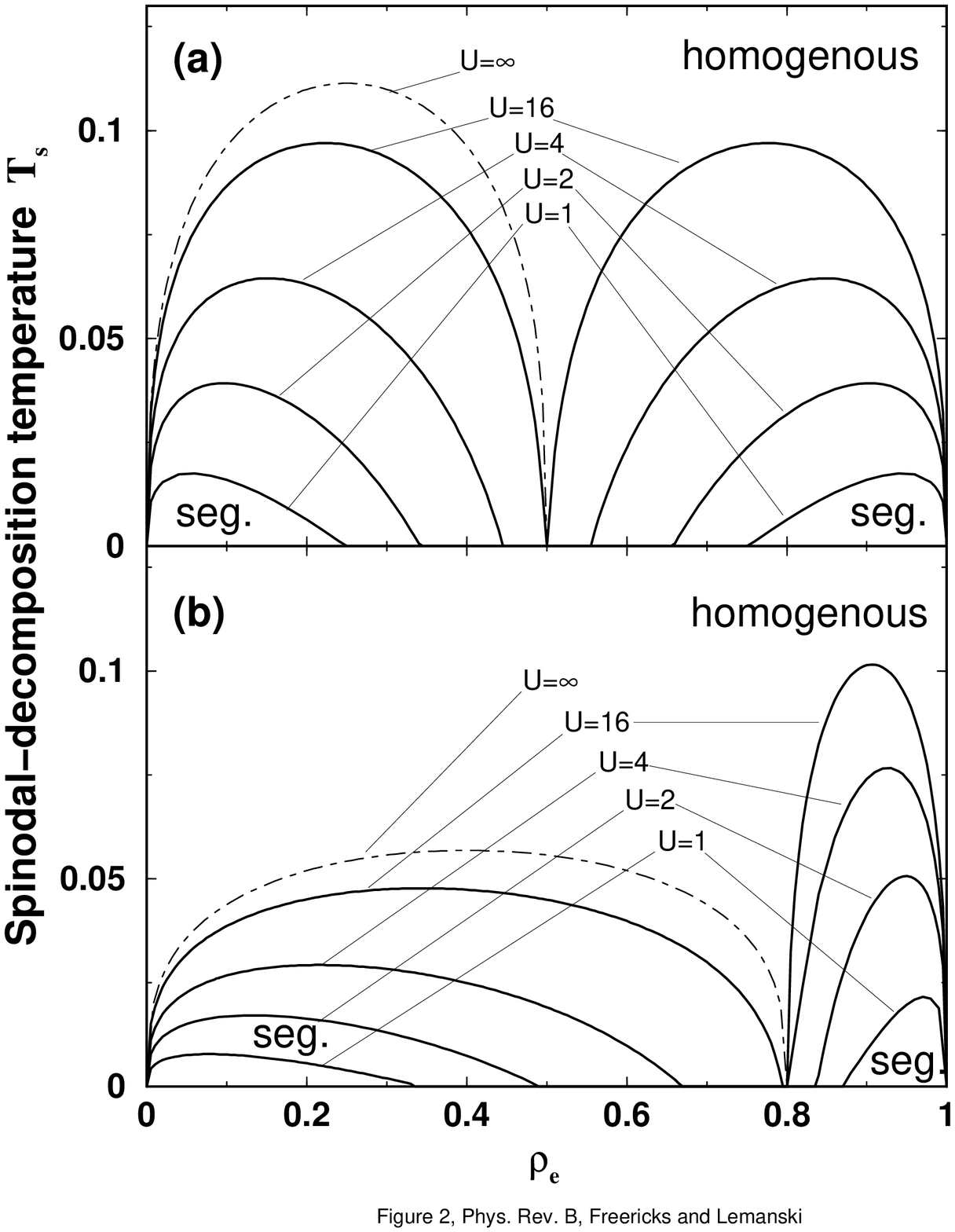}}
\caption{Spinodal decomposition temperature for the segregated phase on
the Bethe lattice as a function
of $U$: (a) the case $\rho_i=0.5$ and (b) the case $\rho_i=0.2$.  The
chain-dashed line corresponds to $U=\infty$ where only the lower
branch is relevant.}
\end{figure}

\begin{figure}[t]
\centerline{
\epsfxsize=5.5in \epsffile{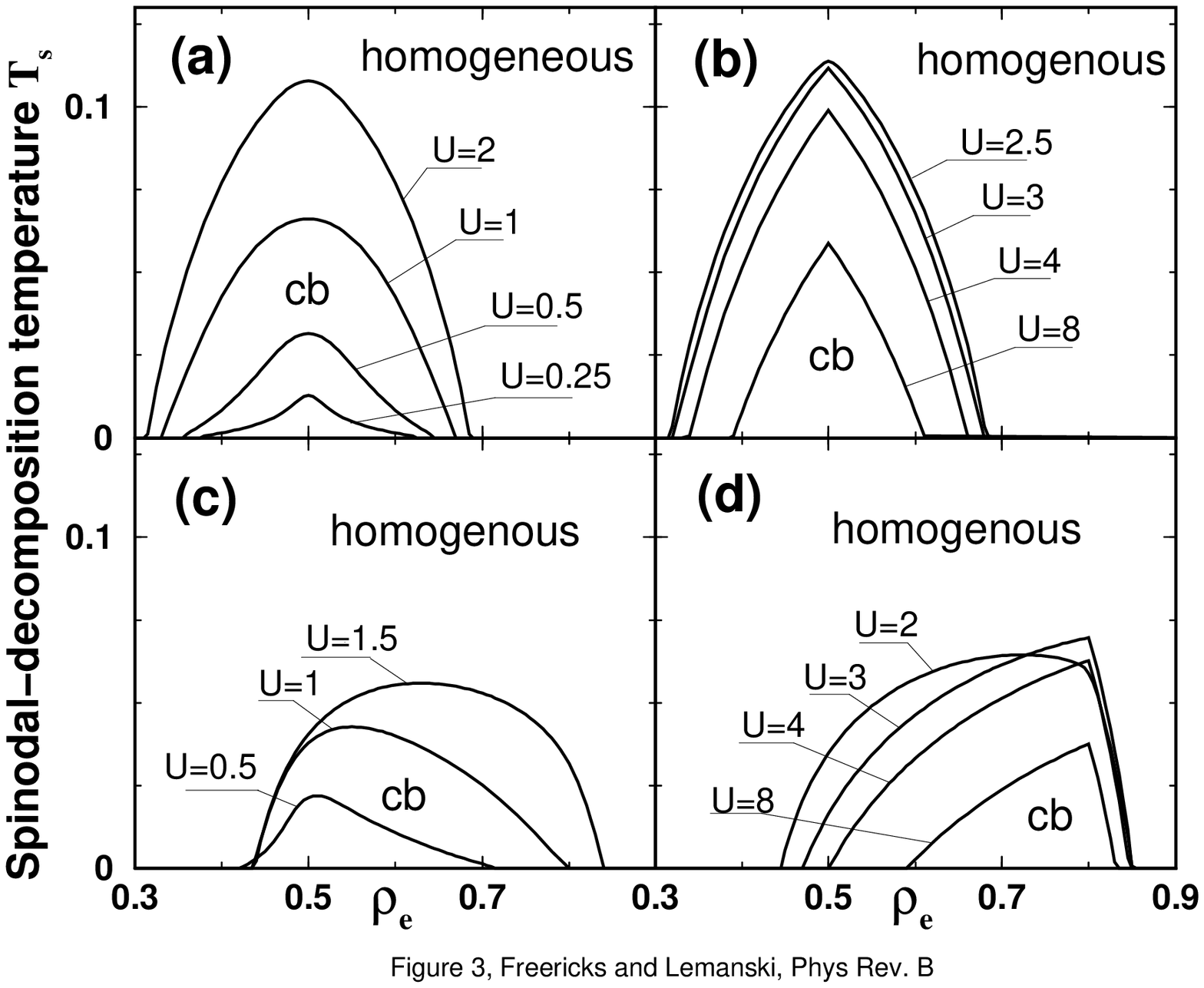}}
\caption{Phase diagram for the chessboard (cb) charge-density-wave phase on
the Bethe lattice: (a) $\rho_i=0.5$ and small $U$, where the curve is smooth;
(b) $\rho_i=0.5$ and large $U$, where the curve develops a kink at $\rho_e=0.5$;
(c) $\rho_i=0.2$ and small $U$, where the curve is smooth; and (d)
$\rho_i=0.2$ and large $U$, where the curve develops a kink at $\rho_e=0.8$.}
\end{figure}

\begin{figure}[t]
\centerline{
\epsfxsize=3.5in \epsffile{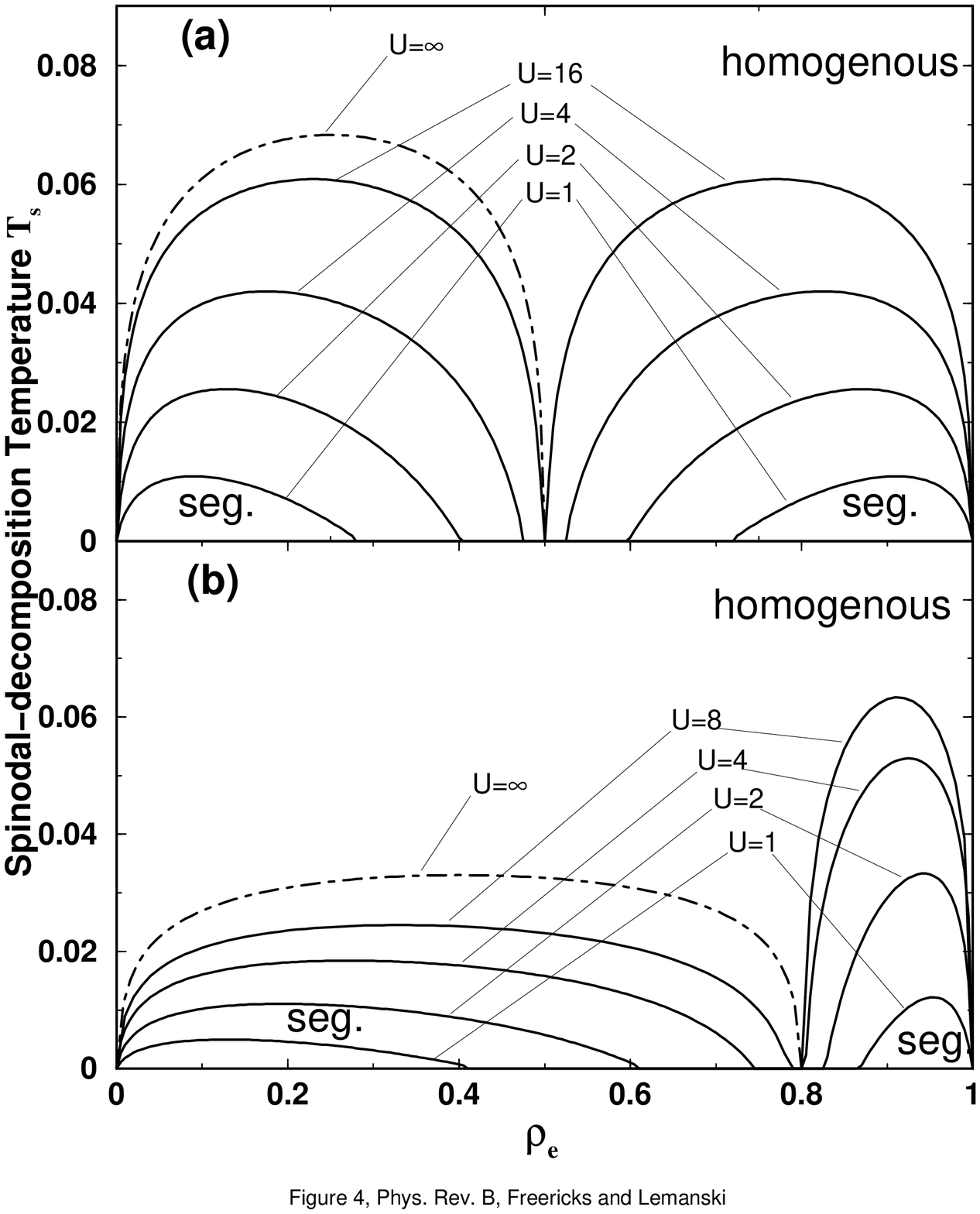}}
\caption{Spinodal decomposition temperature for the hypercubic
lattice as a function
of $U$: (a) the case $\rho_i=0.5$ and (b) the case $\rho_i=0.2$.  The
chain-dashed line corresponds to $U=\infty$ where only the lower
branch is relevant. Note the similarity with Figure 2.}
\end{figure}

\begin{figure}[t]
\centerline{
\epsfxsize=5.5in \epsffile{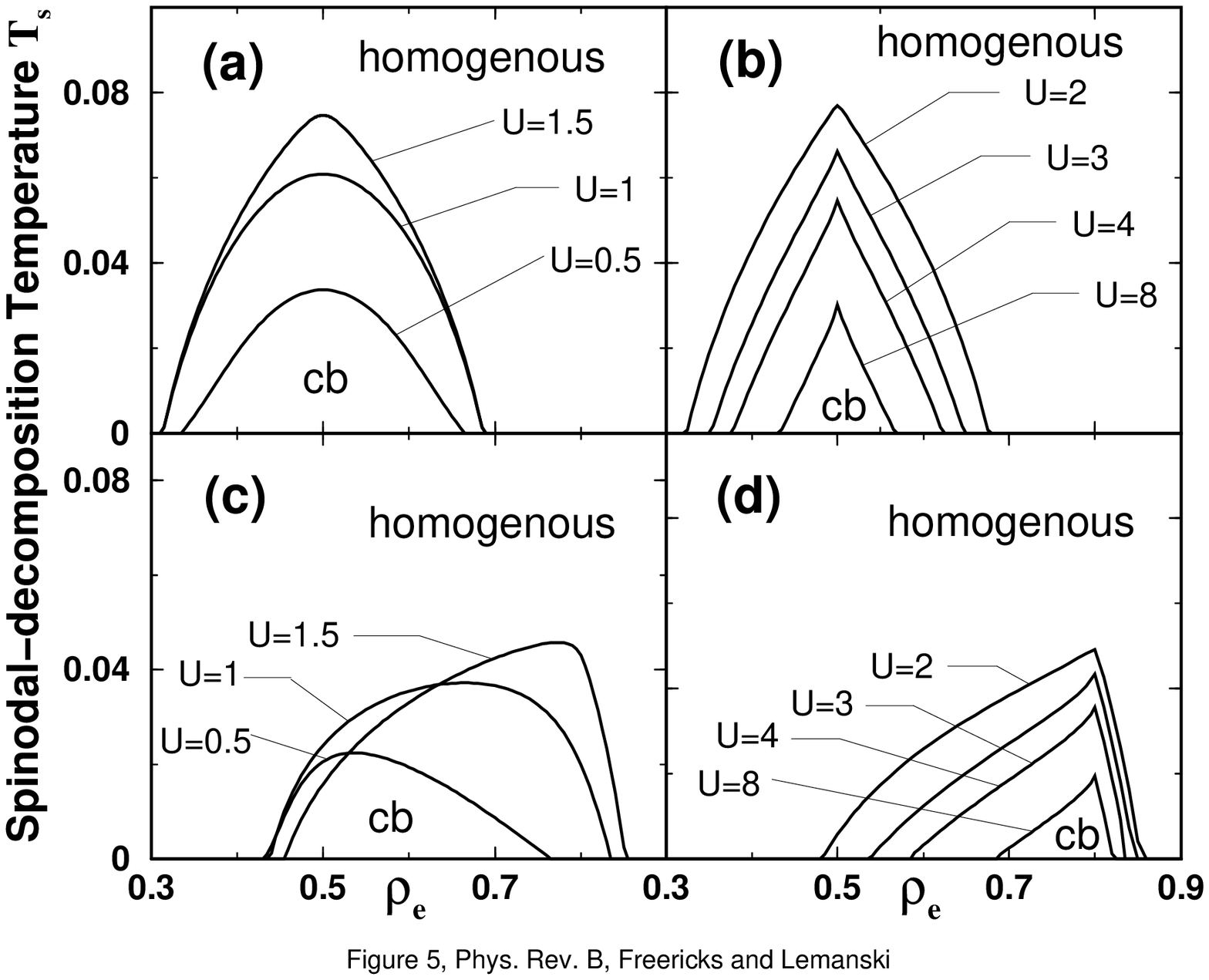}}
\caption{Phase diagram for the chessboard charge-density-wave phase on
the hypercubic lattice: (a) $\rho_i=0.5$ and small $U$, where the curve is 
smooth; (b) $\rho_i=0.5$ and large $U$, where the curve develops a kink at 
$\rho_e=0.5$;
(c) $\rho_i=0.2$ and small $U$, where the curve is smooth; and (d)
$\rho_i=0.2$ and large $U$, where the curve develops a kink at $\rho_e=0.8$.
Note how the only difference with Figure 3 is that the kinks are more strongly
developed here.}
\end{figure}

\begin{figure}[t]
\centerline{
\epsfxsize=3.5in \epsffile{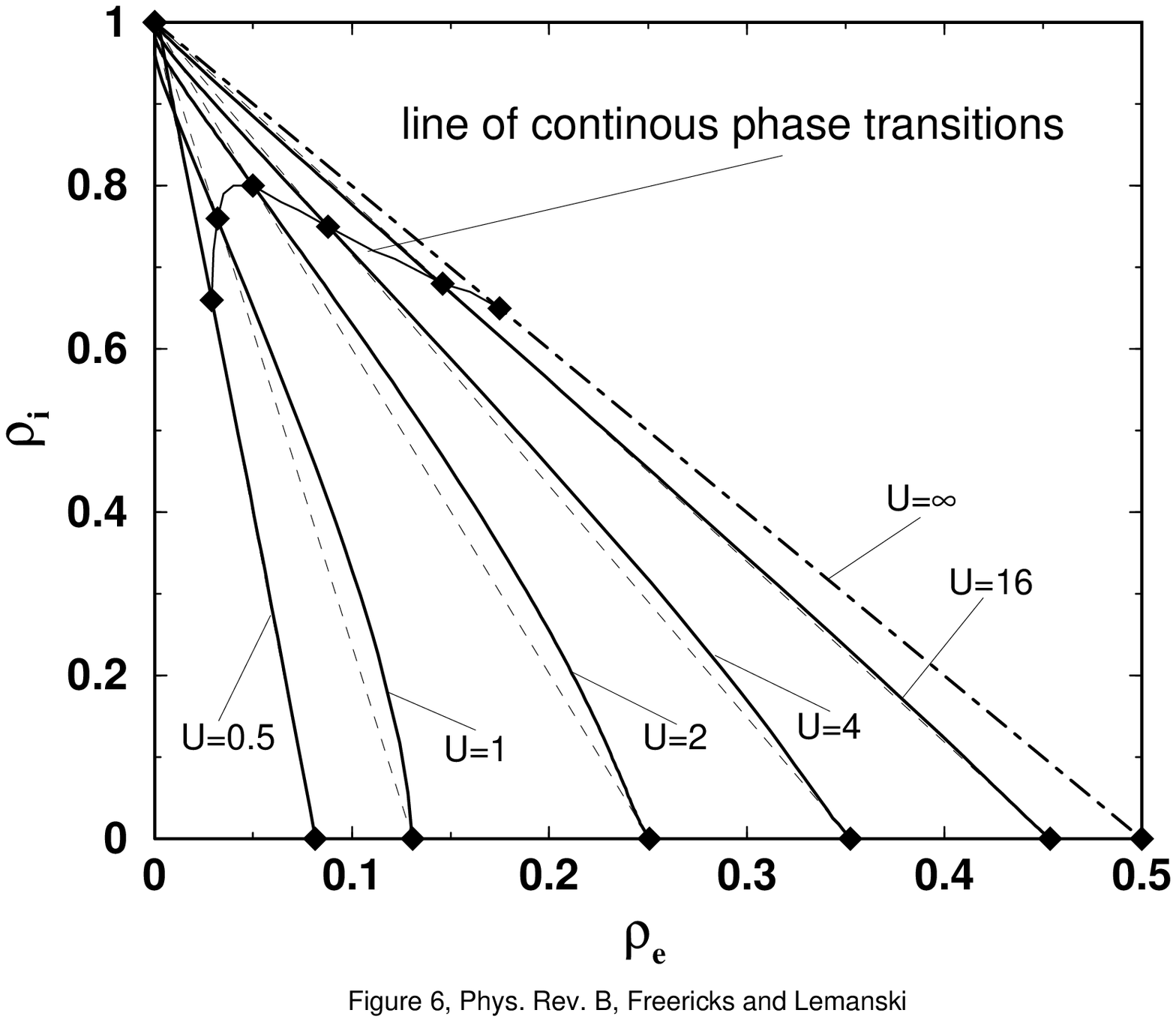}}
\caption{Projections of the segregation phase diagram onto the 
$\rho_e$--$\rho_i$ plane for the Bethe 
lattice.  The solid diamonds connected by the solid curve near
$\rho_i=0.7$ denote the homogeneous densities where
the spinodal-decomposition temperature is a maximum for a given value of $U$.
The solid lines are the values of the densities at various temperatures,
and the dashed lines are straight-line guides to the eye.  The chain-dashed
line is the result with $U=\infty$.}
\end{figure}

\begin{figure}[t]
\centerline{
\epsfxsize=3.5in \epsffile{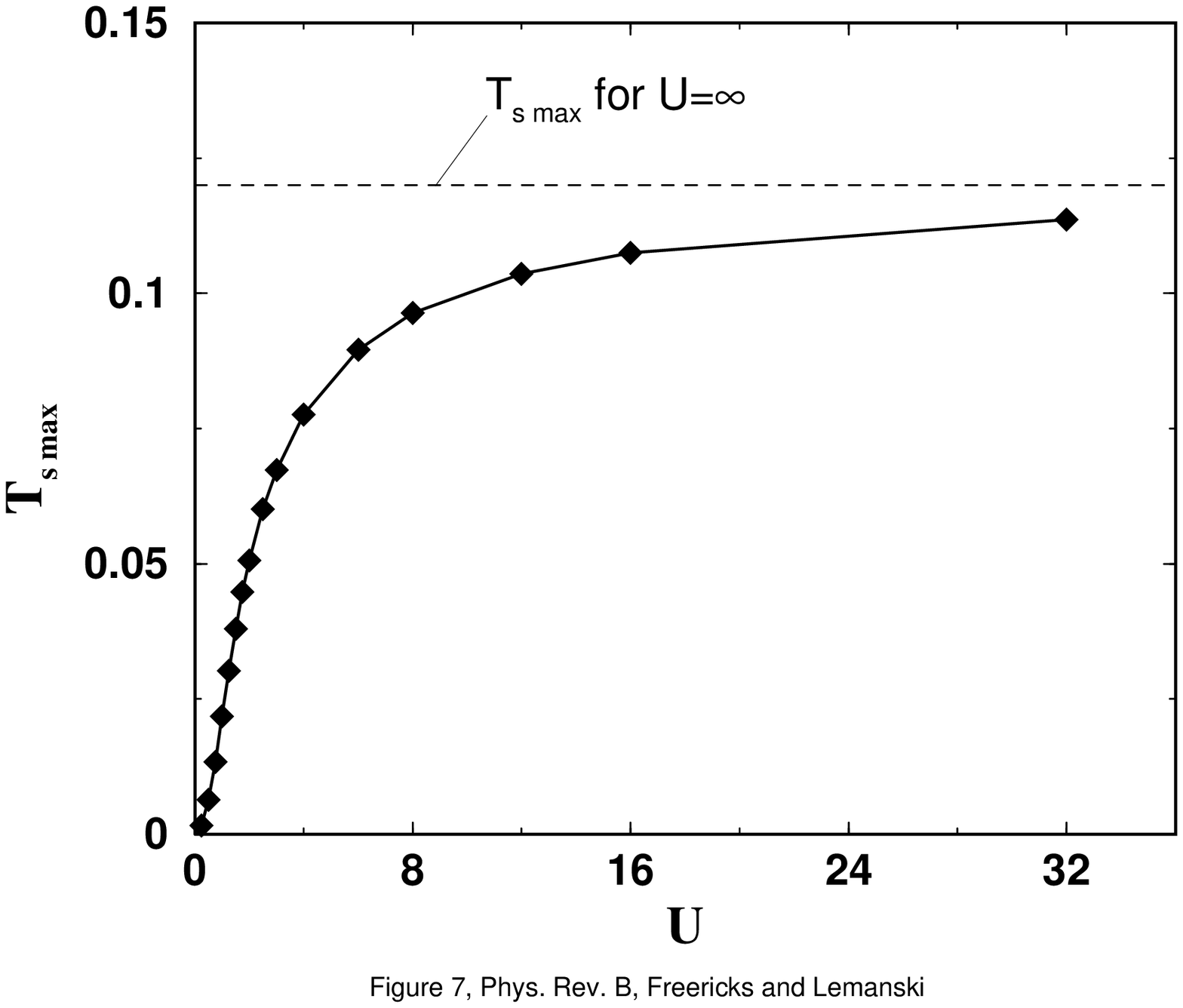}}
\caption{Maximal spinodal-decomposition temperature on the
Bethe lattice plotted as a function of $U$.}
\end{figure}

\begin{figure}[t]
\centerline{
\epsfxsize=3.5in \epsffile{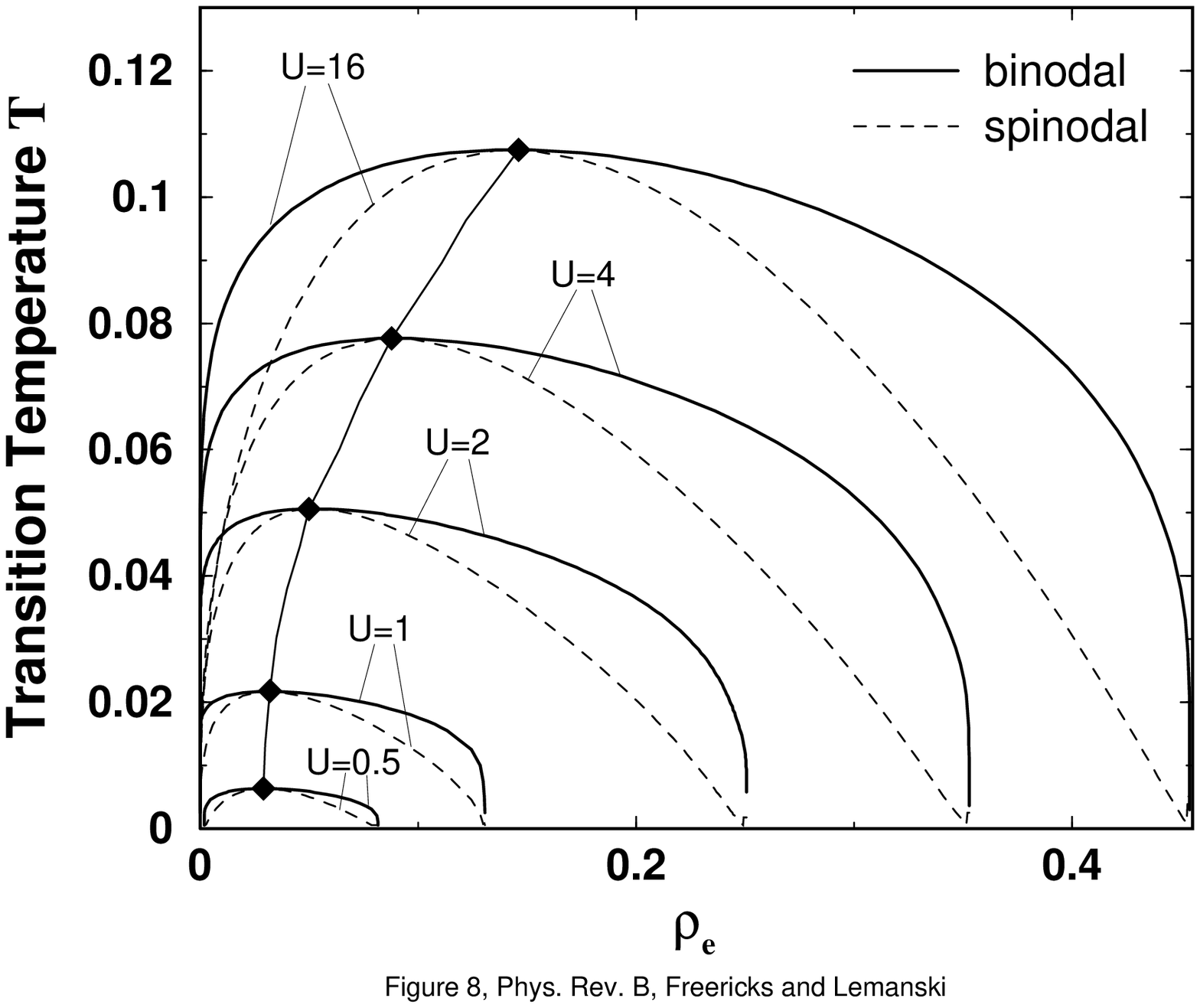}}
\caption{Projection of the segregation phase diagram onto the 
$\rho_e$--$T$ plane for the Bethe 
lattice.  The diamonds denote the homogeneous densities where
the spinodal-decomposition temperature is a maximum for a given value of $U$
(which corresponds to the classical critical point).
The solid lines are the binodal (first-order) transition temperatures,
and the dashed lines are the spinodal-decomposition temperatures.}
\end{figure}

\begin{figure}[t]
\centerline{
\epsfxsize=3.5in \epsffile{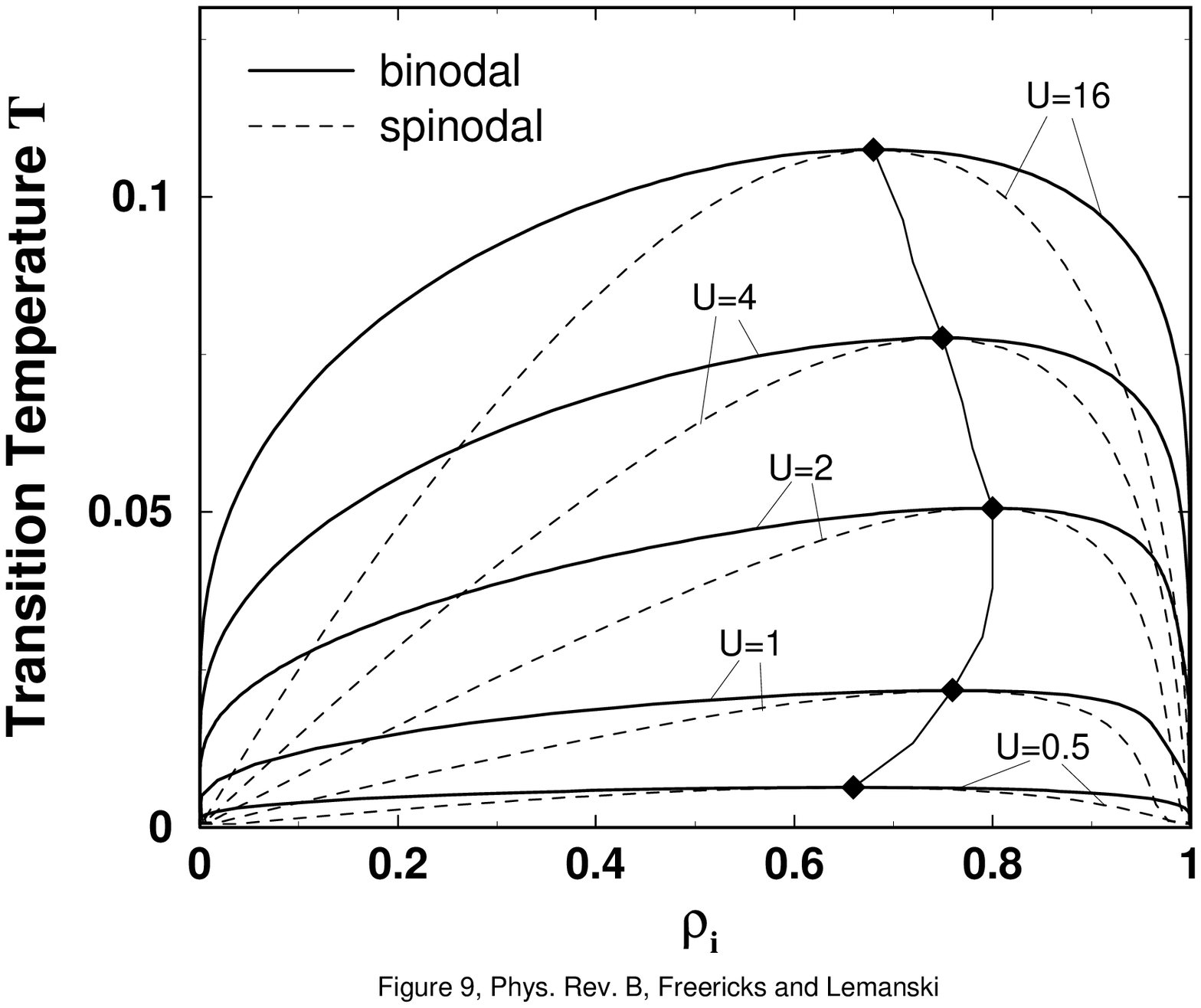}}
\caption{Projection of the segregation phase diagram onto the 
$\rho_i$--$T$ plane for the Bethe 
lattice.  The diamonds denote the homogeneous densities where
the spinodal-decomposition temperature is a maximum for a given value of $U$.
The solid lines are the binodal (first-order) transition temperatures,
and the dashed lines are the spinodal-decomposition temperatures.  Note
how the maximal ion density is not monotonic in $U$.}
\end{figure}

\begin{figure}[t]
\centerline{
\epsfxsize=3.5in \epsffile{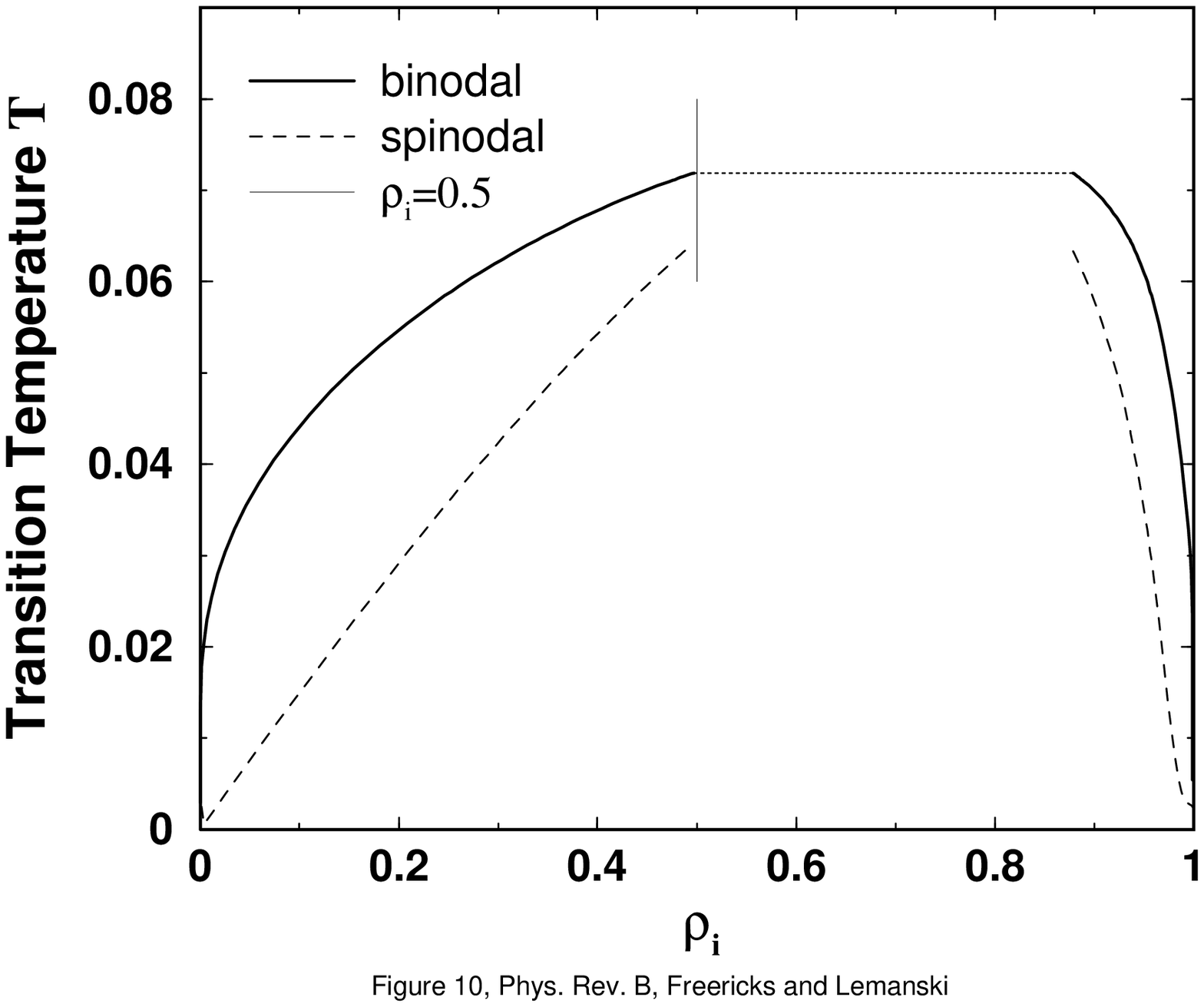}}
\caption{Projection of the segregation phase diagram onto the 
$\rho_i$--$T$ plane for the Bethe 
lattice in a generic discontinuous case ($\rho_e=0.15$ and $\rho_i=0.5$).
The solid line is the binodal (first-order) transition temperature,
and the dashed line is the spinodal-decomposition temperature.  A horizontal
line is included at $\rho_i=0.5$ as a reference.}
\end{figure}

\end{document}